\documentclass[usenatbib]{mn2e}
\usepackage{graphicx}

\begin{document}

\title[Accretion in Short Period Algols]{Three-Dimensional Hydrodynamic Simulations of Accretion in Short Period Algols}

\author[E. Raymer]{Eric Raymer,$^1$ \\ $^1$Department of Physics, North Carolina State University, Raleigh, NC 27695-8202}

\maketitle
\begin{abstract}

Recent observations have shown that the direct-impact Algol systems U CrB and RS Vul possess gas located outside of the orbital plane, including a tilted accretion disc in U CrB. Observations of circumstellar gas surrounding the mass donor in RS Vul suggest magnetic effects could be responsible for deflecting the accretion stream out of the orbital plane, resulting in a tilted disc. To determine whether a tilted disc is possible due to a deflected stream, we use three-dimensional hydrodynamic simulations of the mass transfer process in RS Vul. By deflecting the stream $45^{\circ}$ out of the orbital plane and boosting the magnitude of the stream's velocity to Mach 30, we mimic the effects of magnetic activity near L1. We find that the modified stream parameters change the direct-impact nature of the system. The stream misses the surface of the star, and a slightly warped accretion disc forms with no more than $3^{\circ}$ of disc tilt. The stream-disc interaction for the deflected stream forces a large degree of material above the orbital plane, increasing the out-of-plane flow drastically. Plotting the H$\alpha$ emissivity in velocity space allows us to compare our results with tomographic observations. Deflecting and boosting the stream increases the emissivity in each $v_z$ slice of the out-of-plane flow by at least three and up to eight orders of magnitude compared to the undeflected case. We conclude that a deflected stream is a viable mechanism for producing the strong out-of-plane flows seen in the tomographic images of U CrB and RS Vul.

\end{abstract}


\defcitealias{agafonov06}{Agafonov, Richards \& Sharova (2006)}
\defcitealias{agafonov09}{Agafonov, Sharova \& Richards, 2009}
\defcitealias{blondin95}{Blondin, Richards \& Malinowski, 1995}
\defcitealias{richards10}{Richards, Sharova \& Agafonov, 2010}

\section{Introduction}

The recent use of three-dimensional Doppler tomography to investigate the Algol binaries U CrB \citepalias{agafonov09} and RS Vul \citepalias{richards10} has unveiled a significant amount of detail regarding the accretion process in these systems. As well as confirming the presence of a number of features seen in two-dimensional tomographic data, the new data have shown an unexpectedly large amount of gas located outside of the orbital plane in both systems. U CrB also displays a transient accretion disc inclined with respect to the orbital plane, which is not expected if mass transfer occurs as a result of gravitational and centrifugal forces alone. The companions in short-period Algols are rapidly rotating late-type stars that are likely to be magnetically active.  Magnetic behavior occuring at or near the first Lagrange point (L1) may offer an explanation for the out-of-plane material.

The notion that large-scale magnetic effects may drastically influence the morphology and time-evolution of the accretion processes is evidenced by the observations of \citet{richards10}, who noted the similarity in the out-of-plane velocities located at L1 and magnetically active regions on the donor star in RS Vul. Moreover, they found evidence of a loop prominence on the cool star close to the L1 point in the 3D tomogram of RS Vul, which suggests that the magnetic field may have deflected the gas stream relative to the central plane. Such a deflection would cause gas to be introduced into the system with additional (non-z) angular momentum that could then propagate toward the primary and persist as a tilted disc. This deflection could be attributed to a variety of mechanisms. Starspots on the donor in the prototype Algol system $\beta$ Persei were observed by \citet{olson81}, and coronal ejections in eight other Algol systems, including RS Vul, have been detected by \citet{white83}. More recently, a large-scale persistent coronal loop in Algol was fully resolved by \citet{peterson10}, suggesting the possibility of an extended magnetic field structure between the two stars. While the likelihood of magnetic activity in U CrB and RS Vul is high, the lack of information presently available regarding the exact magnetic structure of these systems makes it difficult to determine the degree to which magnetic effects are responsible for the transport of gas out of the orbital plane.

Further complicating the mass transfer process in Algols is the fact that their short orbital period ($< 5-6$ days) combined with the close orbit and relatively large radius of the primary results in the accretion stream directly impacting or grazing the surface of the accreting star. Direct-impact systems display a permanent or transient Keplerian disc and sometimes a partial disc or localized region characterized by an asymmetric mass distribution and sub-Keplerian velocities \citep{richards92solo, richards99}. In these systems, the localized region is associated with gas that has circled the accreting star and has slowed down because of the interaction with the incoming gas stream \citep{richards92solo}. Gas is expected to leave the orbital plane due to interactions at the site where the stream impacts the surface of the primary, the stream-disc interaction, and the localized region \citep{richards04}.

Both U CrB and RS Vul are short-period binaries, and both display a number of H$\alpha$ emission features characteristic of direct-impact systems. U CrB has been classified as an alternating system, since it has been observed in both a streamlike accretion state and a disclike state \citep{richards99}. The stream state is characterized by strong emission along the predicted ballistic trajectory of the accretion stream against a lack of circumprimary emission. In the disc mode, emission from the accretion disc is the most prominent feature. The mechanism responsible for the transition between these two states has yet to be explained, although hydrodynamic simulations of $\beta$ Per \citepalias{blondin95} showed that the morphology of the accretion flow depended strongly on the density of the inflowing gas. The hydrodynamic simulations of \citet{richards98} also measured a large relative increase in the H$\alpha$ emissivity of the stream relative to the disc when stream density was varied. The similarity of the stellar components and orbital parameters of RS Vul to those of U CrB (RS Vul = B5V + G1 III-IV, P=4.48 days; U CrB = B6V + G0 III-IV, P=3.45 days) makes it likely that a disc mode is possible in RS Vul, and it has been classified as an alternating Algol by \citet{richards99} despite having not yet been observed in a disc mode.

Three-dimensional tomographic imaging of U CrB has allowed for a more thorough analysis of the accretion features by enabling the creation of 3D velocity space plots of the H$\alpha$ emission sources. Mapping this data to Cartesian coordinates is not possible due to the inability to resolve the exact Cartesian structure of the system, but the theoretical analysis of key features of the system provides reference points from which the gas dynamics of the system can be inferred. Features commonly used as landmarks include the motion of the stellar components, the ballistic trajectory of the accretion stream, and the locus of the accretion disc as defined by the Keplerian velocities at the surface of the primary and its outermost Roche surface.

The first 3D tomographic analysis of U CrB was performed by \citetalias{agafonov06}, and showed the system in a disc mode. An additional series of H$\alpha$ profiles were later analyzed in three dimensions and displayed a stream mode \citep{agafonov09}. Notable features of these two modes include: (1) A gas stream transferring mass from the donor to the primary star with velocities as high as 450 km s$^{-1}$ in the orbital plane and negligible z-velocities. (2) Circumprimary emission associated with an accretion disc that displays both positive and negative z-velocities ($v_z$ = $\pm 420$ km s$^{-1}$). The different signs of the z-velocities on opposite sides of the primary star suggest that the disc may be inclined with respect to the orbital plane. (3) A stream-disc interaction region, or ``hotspot'' ($v_z$ = +180 km s$^{-1}$ to +540 km s$^{-1}$), as well as a star-stream impact site ($v_z$ = -300 km s$^{-1}$ to +60 km s$^{-1}$). (4) A localized region ($v_z$ = +200 km s$^{-1}$ to +540 km s$^{-1}$). In contrast to the two accretion modes displayed by U CrB, RS Vul has only been observed in a stream state with a much stronger out-of-plane flow. A three-dimensional tomographic analysis of RS Vul was performed by \citet{richards10}, and displays many of the same features as U CrB. Major differences from U CrB include: a truncated gas stream ($v_z = -85$ km s$^{-1}$) with no indication of a star-stream hotspot, asymmetric circumprimary emission with no well-defined disc, a localized region diametrically opposite the circumprimary emission with opposite z-velocities (suggesting that the gas orbiting the accreting star is flowing along a tilted path with $v_z = +15$ to $-85$ km s$^{-1}$), and emission from a magnetically active region on the donor that suggests the presence of a loop prominence ($v_z = -150$ to $150$ km s$^{-1}$). 

The fact that the H$\alpha$ emission at L1 possesses z-velocities equal to those of magnetically active regions on the donor strongly suggests that the stream is magnetically deflected at or near L1. The relatively high velocities associated with the stream also suggest that the stream velocity may be boosted to many times the speed that would occur via Roche lobe overflow alone. Our goal was to determine whether a deflected stream is responsible for introducing a sufficient amount of tilt to the accretion disc, and to what extent the deflection influences the presence or absence of other out-of-plane accretion features. To investigate this we performed hydrodynamic simulations of mass transfer in RS Vul in which magnetic effects were mimicked by manually deflecting and boosting the stream at L1. The resulting steady-state solutions were analyzed to determine the effect of the deflection on the morphology of the circumstellar material, the presence of out-of-plane flow, and the presence or absence of the accretion features noted in the tomographic data. The hydrodynamic model we used is described in section 2, results from runs with undeflected and deflected streams are given in section 3, an analysis of the tilt of the disc with comparisons to analytical estimates is given in section 4, and our results summarized in section 5.

\section{Computational Methods}

\subsection{Grid parameters}

Our simulations were performed with the three-dimensional hydrodynamics code VH-1, which was developed by the Numerical Astrophysics Group at the Virginia Institute for Theoretical Astrophysics. VH-1 solves the Euler equations for the flow of an inviscid fluid characterized by an adiabatic index $\gamma$. To solve the system, VH-1 utilizes operator splitting to reduce the grid into a number of 1-D segments. The fluid equations are solved along each segment using the Piecewise Parabolic Method, an explicit high-order Godunov method employed on an Eulerian (fixed) grid \citep{colella84}. The PPM can be used for both 2D and 3D geometries without necessitating significant changes to the algorithm other than the extension from 2D polar to 3D spherical coordinates.

We adapted VH-1 to model our system in spherical coordinates on a corotating grid located in the reference frame of the binary system and centered upon the primary star.  The inner boundary of the grid coincides with the surface of the primary at a radius of $R_1$, and the outer boundary extends to twice the distance from the centre of the primary to L1 ($\sim 6.2 R_1$). Since the system does not vary widely in scale between the two stars, we space our zones equally in the radial direction. The parameters of U CrB and RS Vul are given in Table 1, and were taken from \citet{agafonov09} and \citet{richards10}. We note that while we use the parameters corresponding to RS Vul in our simulations, both the orbital parameters and stellar components of U CrB are similar to those of RS Vul. This enables us to interpret our results in the context of either system. 

\begin{figure}
\includegraphics[width=84mm]{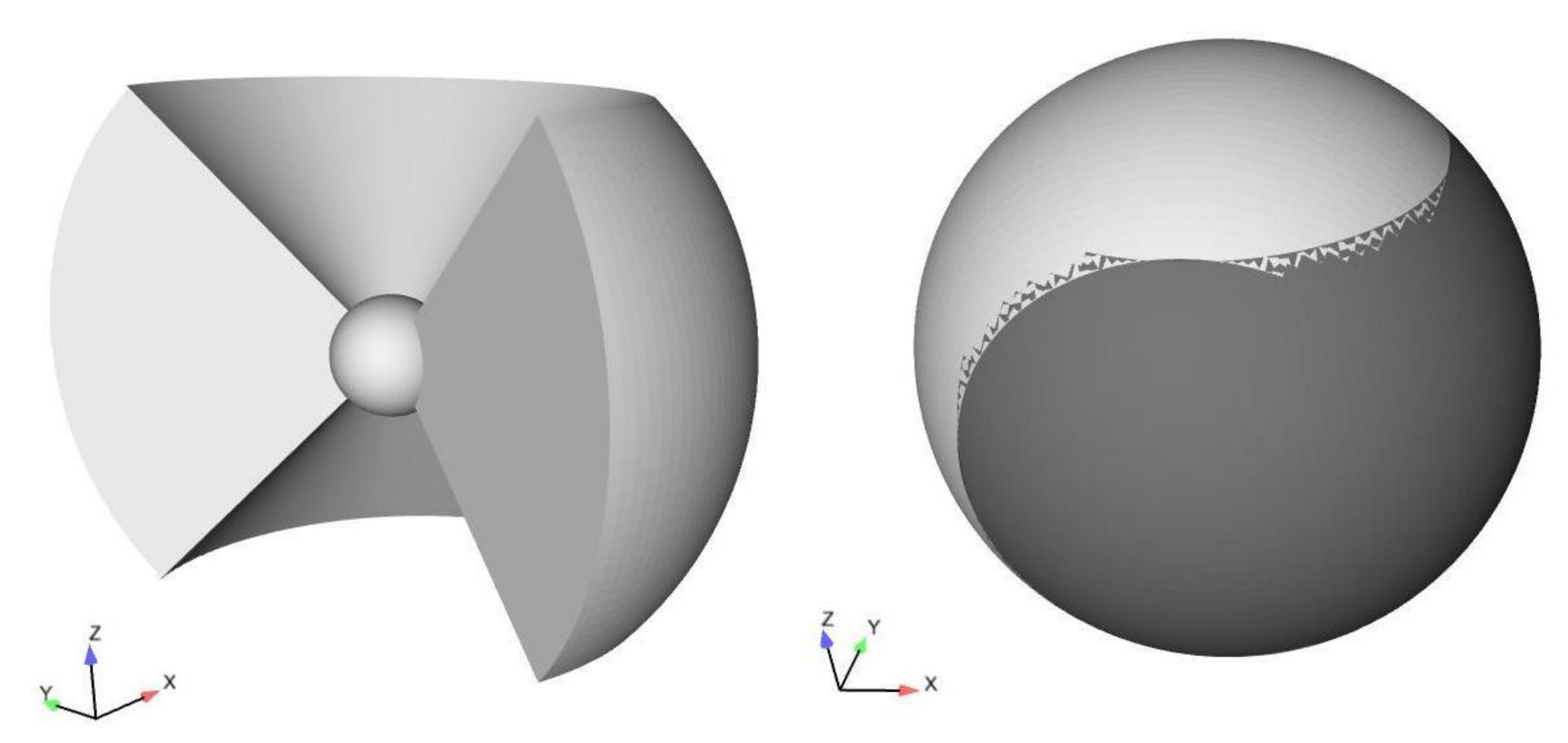}  
\caption{Yin-yang grid geometry. The figure on the left shows the range of one grid, and the figure on the right shows the overlap between the two grids.}
\label{fig:yinyang}
\end{figure}

To avoid grid convergence and the coordinate singularities intrinsic to spherical coordinates, we employ a yin-yang overset grid \citep{kageyama04}. The yin-yang grid consists of two separate spherical polar grids oriented 90$^\circ$ with respect to each other, so that the polar region of one grid aligns with the equatorial region of the other. Each grid extends $\pm \pi/4$ azimuthally, and $\pm 3\pi/4$ in the polar direction (Figure \ref{fig:yinyang}). This grid configuration allows for reasonable limits on the Courant number and ensures that our runtimes do not become excessive. As a result, we are able to extend our outer boundary to a sufficiently large radius without imposing limits on the the azimuthal extent of the grid. We define our coordinates such that the positive x-axis of the yin grid lies along the binary axis and points toward the secondary. The positive y-axis points in the direction of motion of the secondary, and the z-axis is perpendicular to the orbital plane. Each piece of the grid has dimensions of $288 \times 144 \times 432$, which results in an angular resolution of $\Delta\phi \sim 0.625^\circ$ and $\Delta r/R_1$ = 0.018 at the inner boundary.

Since the inner boundary corresponds to the surface of the accreting star, we must choose a boundary condition that allows gas to accrete while still adequately modeling the effects of the stellar atmosphere. The fact that the stream penetrates only a small portion of the stellar atmosphere \citep[][$\sim 10^{-6}R_1$ for Algol, ]{richards92solo} allows us to approximate the surface of the primary as a solid surface. Interactions between the atmosphere and the stream could have a non-negligible effect on the structure of the circumstellar gas (especially in the case of a rotating and precessing primary), but for the purposes of this paper we neglect the structure of the atmosphere. Instead, we employ a partially reflecting boundary condition that sets the radial velocity of the gas at the innermost radial zone to zero every timestep. This approximates the gas encountering a shock at the surface of the star and accreting shortly afterward. At the outer boundary, gas should be allowed to freely exit the grid, so we implement a modified zero-gradient boundary condition with the added requirement that gas may not be allowed to flow inward once it has passed off the grid.

\begin{table*}
\begin{minipage}{126mm}
\caption{Parameters for U CrB and RS Vul}
\label{symbols}
\begin{tabular}{lll}
\hline
Parameter & U CrB & RS Vul \\
\hline
Mass of Primary, $M_1$ & $4.80 M_{\odot}$ & $6.59 M_{\odot}$ \\
Mass of Secondary, $M_2$ & $1.40 M_{\odot} $ & $1.76 M_{\odot}$ \\
Effective Temperature of Primary, $T_1$ & $14000$ K & $16500$ K \\
Effective Temperature of Secondary, $T_2$ & $5250$ K & $6000$ K \\
Radius of Primary, $R_1$ & $3.0 R_{\odot}$ & $4.7 R_{\odot}$ \\
Radius of Secondary, $R_2$ & $4.6 R_{\odot} $ & $5.8 R_{\odot}$ \\
Binary Separation, $a$ & $1.23 \times 10^{12}$ cm & $1.61 \times 10^{12}$ cm \\
Orbital Period, $P_{orb}$ & $3.452$ days & $4.478$ days \\
\hline
\end{tabular}
\end{minipage}
\end{table*}

\subsection{Radiative cooling and stream density}

The 2D simulations of Algol by \citet{blondin95} found that the morphology of the circumprimary flow depended strongly on the initial stream density due to the effects of radiative cooling. The stream density's influence on the flow is seen primarily in the stream-disc interaction. A denser stream will cool more rapidly at the shock, negating the effects of heating within the shock. This results in a flow that behaves similarly to a compressible isothermal gas. A lower-density stream will not cool to the same extent, and will behave adiabatically.

To investigate the relative effect of cooling in U CrB and RS Vul, we compare the cooling time at the shock to a characteristic dynamical time. The dynamical time can be found using the gas speed at the outer edge of the disc where the stream-disc shock is located. Assuming a Keplerian velocity at the outer edge of the disc ($\approx 350$ km s$^-1$) allows us to calculate a dynamical time of:

\begin{equation}
t_d = \frac{GM_1}{v_{Kep}^3}
\end{equation}

In both U CrB and RS Vul, the dynamic timescale is calculated to be $0.04 P_{orb}$.

If a cooling function is known, we can calculate the characteristic cooling time at the shock to be

\begin{equation}
t_c = \frac{kT}{n_s\Lambda(T)}
\end{equation}

Here, $n_s$ is the electron number density in the post-shock flow, and $\Lambda$ is the cooling function evaluated at the temperature of the shocked gas, $T$. We calculate $\Lambda(T)$ with the cooling function for a low-density plasma as given by \citet{cox71}. From the Rankine-Hugoniot jump conditions, we can calculate the post-shock temperature to be $T=3\tilde{m}v^2_{sh}/16k$, where $\tilde{m}$ is the average mass per particle. We assume a gas comprised largely of ionized Hydrogen, in which case $\tilde{m} \sim m_H/2$. We assume the velocity of the shocked gas, $v_{sh}$ to be on the order of the Keplerian velocity at the shock site. For both U CrB and RS Vul, we calculate $T_{sh}$ to be approximately $1 \times 10^6$ K, which results in a cooling coefficient of $\Lambda = 8 \times 10^{-23}$ ergs cm$^3$ s$^{-1}$. An analytical estimate of the density at the shock site is difficult to obtain, so we note that our simulations show densities in the shock region on the order of $\rho = 0.1 \rho_{L1}$. Converting this density to physical units requires a measurement of the mass transfer rate. Since data on the exact mass transfer rates for U CrB and RS Vul are not available, we use the lower limit of the transfer rate in Algol, $\dot{M} \approx 10^{-11} M_{\odot}$ yr$^{-1}$ \citep{richards92solo}. 

Comparing the characteristic dynamical time to the cooling time, we find that $t_d / t_c = 2.4$. This is a conservative estimate, as we have used a small value of the mass transfer rate to calculate the cooling time. Any increase in the mass transfer rate will serve to decrease the cooling time further, since $t_c$ varies as $1/\dot{M}$. In general, the mass transfer rates in short-period Algols vary widely between systems, and are not likely to remain constant in time. For Algol, $\dot{M}$ could range from $10^{-11}$ to $10^{-9} M_{\odot}$ yr$^{-1}$ \citep{richards92solo}. With these factors taken into account, it is likely that cooling at the shock is sufficiently rapid to counter the heating that occurs in the stream-disc interaction. This rapid cooling ensures that the shock-heated gas will quickly fall to some minimum temperature below $10^4$ K, at which point the recombination of H will diminish the cooling rate. The cooled gas will be heated by the strong photoionizing UV flux from the primary, which is sufficient to maintain the temperature at a value close to $10^4$ K. Taking this information into account, we assume a constant temperature of $11,000$K  in our simulations. This value falls midway between the effective temperatures of the primary and secondary in both U CrB and RS Vul. This allows us to expedite our calculations by employing an isothermal equation of state ($\gamma = 1$). 

\subsection{Initial conditions at L1}

\citet[hereafter LS]{lubow75} conducted an extensive analysis of the gas dynamics at the L1 point, and we use their prescription to determine the initial conditions of the accretion stream. LS parametrized the flow in terms of the ratio of isothermal sound speed to the system's orbital speed: $\varepsilon = c_s / \omega a = (kT/m)^{1/2}$, for which we obtain $\varepsilon=0.035$. This gives a local sound speed of $9.5$ km s$^{-1}$ at L1. LS calculate that the stream will be accelerated to supersonic speeds as it travels through L1, so we assume an unboosted speed of Mach 1. The angle the stream makes with the binary axis can also be calculated using $\varepsilon$, and for RS Vul we obtain $\theta = 20.4^{\circ}$. Assuming an ideal gas, the pressure at L1 should be initialized to a value on the order of $\rho \varepsilon^2$. 

For this work we require control over the direction and magnitude of the gas velocity at L1 in order to manually deflect the stream. To achieve this, we calculate the position of L1 on the grid and hold the stream parameters in the corresponding zones fixed at the values given above. This eliminates the need for the donor star to be modeled hydrodynamically, and we remove it from the grid while retaining its term in the gravitational potential. A small amount of flow into the region corresponding to the secondary occurs, but it remains confined to the volume that the secondary would normally occupy and does not affect the behavior of the accretion flow. In addition to giving us more control over the stream at L1, this strategy has the benefit of allowing for a more economic grid size, since the full secondary need not be present on the grid.

LS calculate the width of the stream at L1 to be on the order of $\varepsilon a$, which requires an angular resolution of at least $\Delta\phi = 0.022$. The angular resolution of our grid is small enough to accomodate this with two zones; each has an angular extent of $\Delta\phi = 0.011$. Symmetry requires the gas stream at L1 to extend equal distances above and below the orbital plane as well as an equal distance to either side of the binary axis. This necessitates the use of four zones to represent L1. In reality the density profile of the gas at L1 is Gaussian, but we lack the resolution to resolve this. Instead we hold the density of the gas fixed to a constant at our L1 zones. Ultimately, our system exhibits behavior similar to that of \citet{blondin95} in that the morphology of the accretion flow is relatively insensitive to the geometry of the zones at L1.

Our simulations do not assume any gas to be initially present around the primary, and the system requires many orbital periods to allow the full formation of a disc. A steady-state solution in which a disc is present and has stopped gaining mass (indicated by $\dot{M}_{acc} = \dot{M}_{L1}$) requires $40 - 50$ P$_{orb}$ according to preliminary low-resolution runs. To expedite the disc formation, we multiply the density in the disc by a constant factor after two orbital periods have elapsed. Since the mass accretion rate is proportional to the density at the inner boundary, this results in the mass accretion rate being boosted to a value closer to that at which it stabilizes. 

\section{Results}

\subsection{Undeflected stream}

\begin{figure*}
\begin{minipage}{126mm}
\includegraphics[width=126mm]{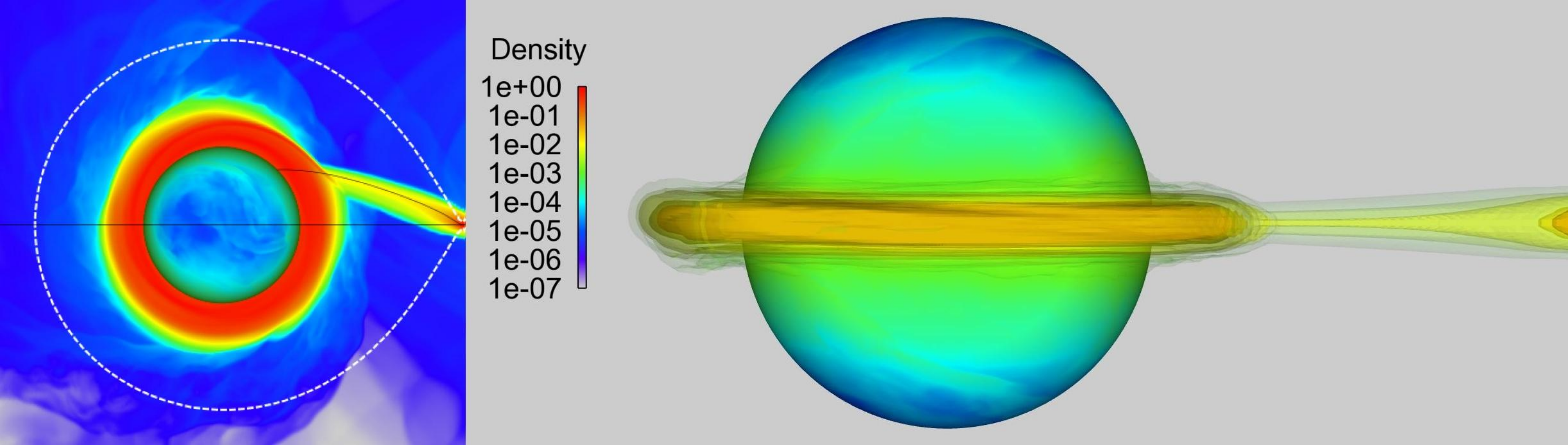}  
\caption{Top and side views of the primary star and and disc for an undeflected stream after $18 P_{orb}$. The density scale is normalized to 1 at L1. The top view depicts the density on a slice through the orbital plane. Density isosurfaces in the side view are taken at $\rho = 1 \times 10^{-1}, 5 \times 10^{-2}, 1 \times 10^{-2}, 5 \times 10^{-3},$ and $1 \times 10^{-3}$ with the transparency increased by a factor of 0.5 for each isosurface.}
\label{fig:notilt_topside}
\end{minipage}
\end{figure*}

To ensure that our model behaved appropriately in the absence of a deflected stream, we performed a fiducial run with the stream left unmodified ($\phi=0^{\circ}$, $v_{stream}=1M$). Figure \ref{fig:notilt_topside} shows the stream following the ballistic trajectory as it falls toward the primary and narrowing as it approaches the surface. The stream impacts the surface after approximately $0.1P_{orb}$. We allow gas to accrete until $2 P_{orb}$, during which a thin accretion disc begins to form. At $2 P_{orb}$, we boost the disc density. The boost instigates some radial density oscillations due to increased pressure along the density discontinuity that occurs at the radius of the boosted region. These oscillations propagate with the flow around the disc, but fade within the next few orbital periods. Following the density boost, the radius of the disc continues to increase for several orbital periods, and a steady state is achieved at roughly $15 P_{orb}$. 

\begin{figure}
\includegraphics[width=84mm]{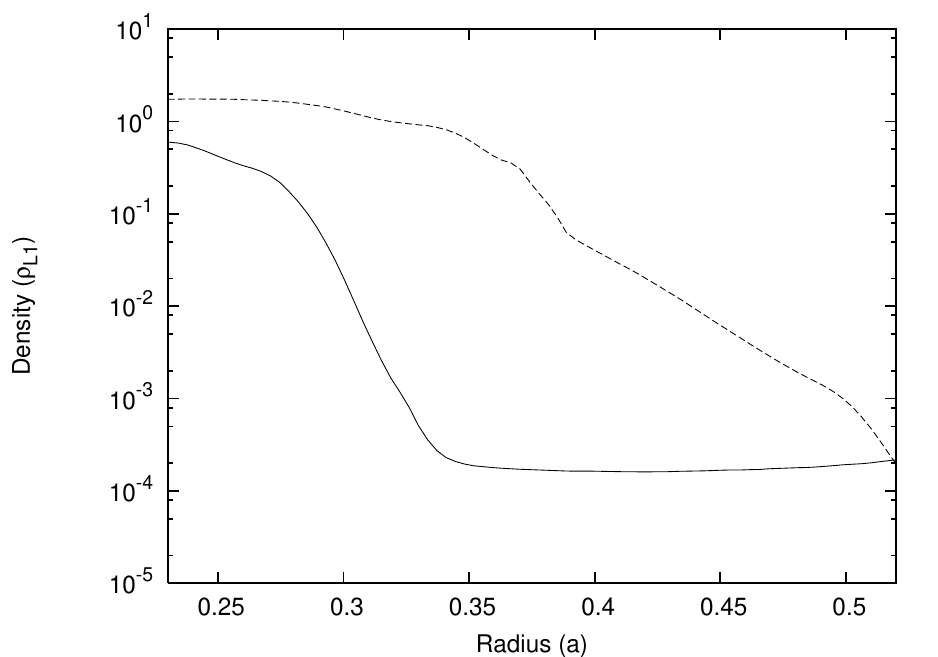}  
\caption{Density vs. radius, averaged over $\phi$ within the orbital plane at $18 P_{orb}$. Densities are normalized to 1 at L1. The solid line depicts the density for the undeflected stream, while the dashed line corresponds to a stream tilted to $45^{\circ}$ and boosted to Mach 30. L1 is located at $0.63a$, and the surface of the primary at $0.20a$.}
\label{fig:density_v_radius}
\end{figure}

When the system has stabilized, we measure the disc to have an outer radius of $0.34a$ (where $a$ is the binary separation as measured from the center of the accreting star). Examining the density profile averaged over $\phi$ (Figure \ref{fig:density_v_radius}) shows that the density of the innermost portion of the disc ($\sim 0.22a$) is of the same order of magnitude as the density at L1 ($\sim 0.63a$). The density decreases sharply to just above $\sim 10^{-4} \rho_{L1}$ at the outer edge of the disc. The disc is symmetric about the orbital plane to approximately $\pm 0.03a$ above and below the plane. Beyond this region the density decreases rapidly, and is measured to be less than $10^{-4}\rho_{L1}$.

The velocities within the disc are oriented largely within the plane. As the gas accelerates from L1, its speed increases from $\sim 10$ km s$^{-1}$ to a final velocity of $\sim 240$ km s$^{-1}$ before encountering the stream-disc shock. (All velocities are measured with respect to a corotating frame.) After crossing the shock and entering the higher-density regions of the disc, the gas assumes velocities which fall within the range of $240-440$ km s$^{-1}$, with small velocities corresponding to the outer edge of the disc, and higher velocities located near the surface of the accretor. The velocities within the disc remain relatively constant in direction and are sub-Keplerian throughout ($60- 80$ percent at the outer and inner edges of the disc, respectively). These outer velocities are less uniform near the localized region and stream-disc interaction site, and reach values up to $300$ km s$^{-1}$ in the vicinity of the stream-disc shock.

\begin{figure}
\includegraphics[width=84mm]{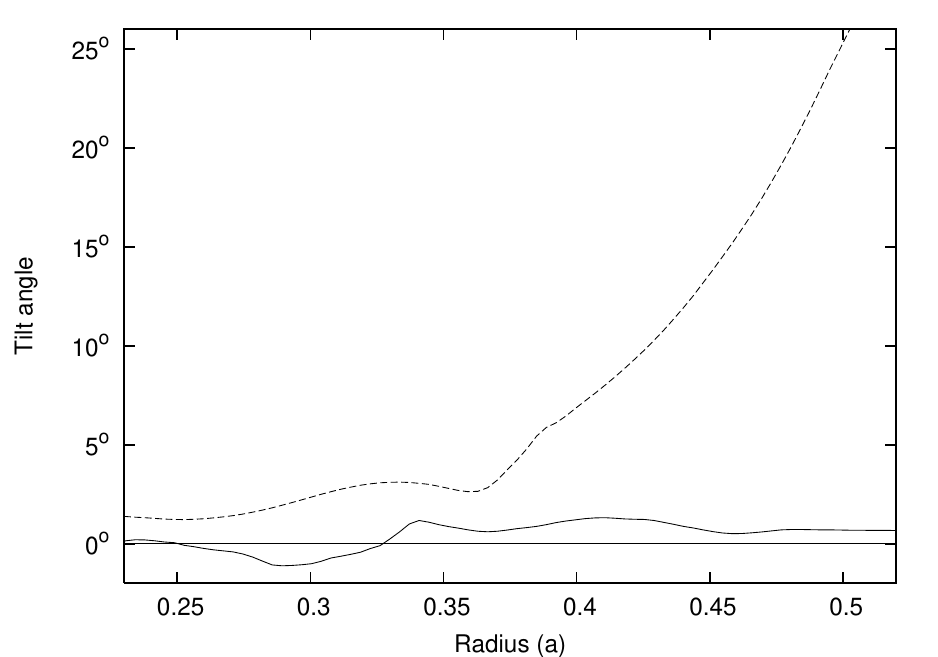}  
\caption{Tilt of the disc vs. radius, calculated at $18 P_{orb}$ by summing the angular momentum within the disc over $\phi$ for each value of $r$. The solid line depicts tilt for the undeflected stream, while the dashed line corresponds to a stream tilted to $45^{\circ}$ and boosted to Mach 30.}
\label{fig:tilt_v_radius}
\end{figure}

To quantify the angle with which the disc is inclined, we integrate the angular momentum present in the disc and measure its angle with respect to the z-axis. We limit our integration to the region occupied by the high-density portions of the disc, using Figure \ref{fig:density_v_radius} as a guideline. The result is summed angular momentum values for each value of $(r, \phi)$ in the orbital plane. Each value of $r$ defines a ring within the disc for which we are able to calculate the tilt. The sign of the tilt is set equal to the sign of the x-component of the net angular momentum. This has the effect of assigning positive tilt angles to rings that are tilted toward the donor, and negative angles for rings that are tilted away. For the undeflected stream, the disc is largely symmetric about the orbital plane and remains essentially untilted over time. Examining the tilt as a function of radius at $18 P_{orb}$ (Figure \ref{fig:tilt_v_radius}) shows that the high-density portions of the disc from the inner boundary to $\sim 0.3a$ only possess tilt on the order of $0.1^{\circ}$. This is a fraction of the angular resolution of by one grid zone. The tilt  varies gradually as the radius increases, and the sign changes three times along the radius, indicating that some warping occurs about the orbital plane. The tilt becomes slightly larger in the low-density portions of the disc and reaches a maximum value of $0.5^{\circ}$. This tilt is smaller than the amount able to be resolved with one zone. The presence of this tilt is likely a result of a comparatively large amount of non-z angular momentum created in the stream-disc interaction through the splashing of low-density gas. Turbulence in the gas flowing along the outer low-density edge of the disc may also contribute to this measurement. 

\begin{figure}
\includegraphics[width=84mm]{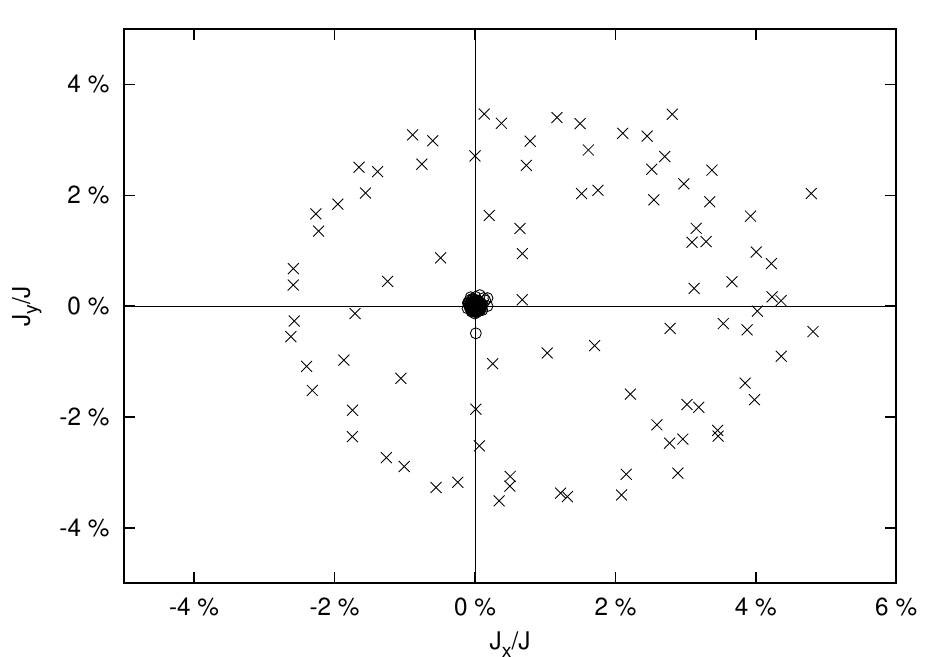}  
\caption{Components of the angular momentum of the disc for the undeflected stream (circles) and the deflected and boosted stream (crosses). Data points were taken at intervals of $\sim 0.2 P_{orb}$. The undeflected stream results in an disc that remains largely in the orbital plane, with minor variations over time that remain symmetric about the origin. The deflected and boosted stream produces a disc biased toward $+J_x$ due to the stream impacting the far side of the disc from below and tilting it toward the donor star.}
\label{fig:orientation}
\end{figure}

The disc's orientation with respect to the orbital plane can be found by examining a plot of the x and y components of the angular momentum. In cartesian space, the projection of the angular momentum vector into the orbital plane points toward the direction that the disc is azimuthally inclined. (This is analogous to treating the disc as a rigid body with a rod through its axis of rotation. The projection of the angular momentum vector into the orbital plane is the direction you would push the rod to tilt the disc.) Figure \ref{fig:orientation} shows a $J_y$ vs. $J_x$ plot for both runs, with $J_x$ and $J_y$ components shown as a fraction of the total angular momentum. The data points for the undeflected stream lie clustered around the origin, with magnitudes of less than $1$ percent of the total angular momentum. This supports the qualitative observation that the disc lies almost entirely in the orbital plane throughout the length of the simulation. The disc appears to precess at a period of $P_{orb}$, which is a by-product of measurements of the angular momentum taken from within a co-rotating reference frame.

\subsection{Deflected stream}

\begin{figure*}
\begin{minipage}{126mm}
\includegraphics[width=126mm]{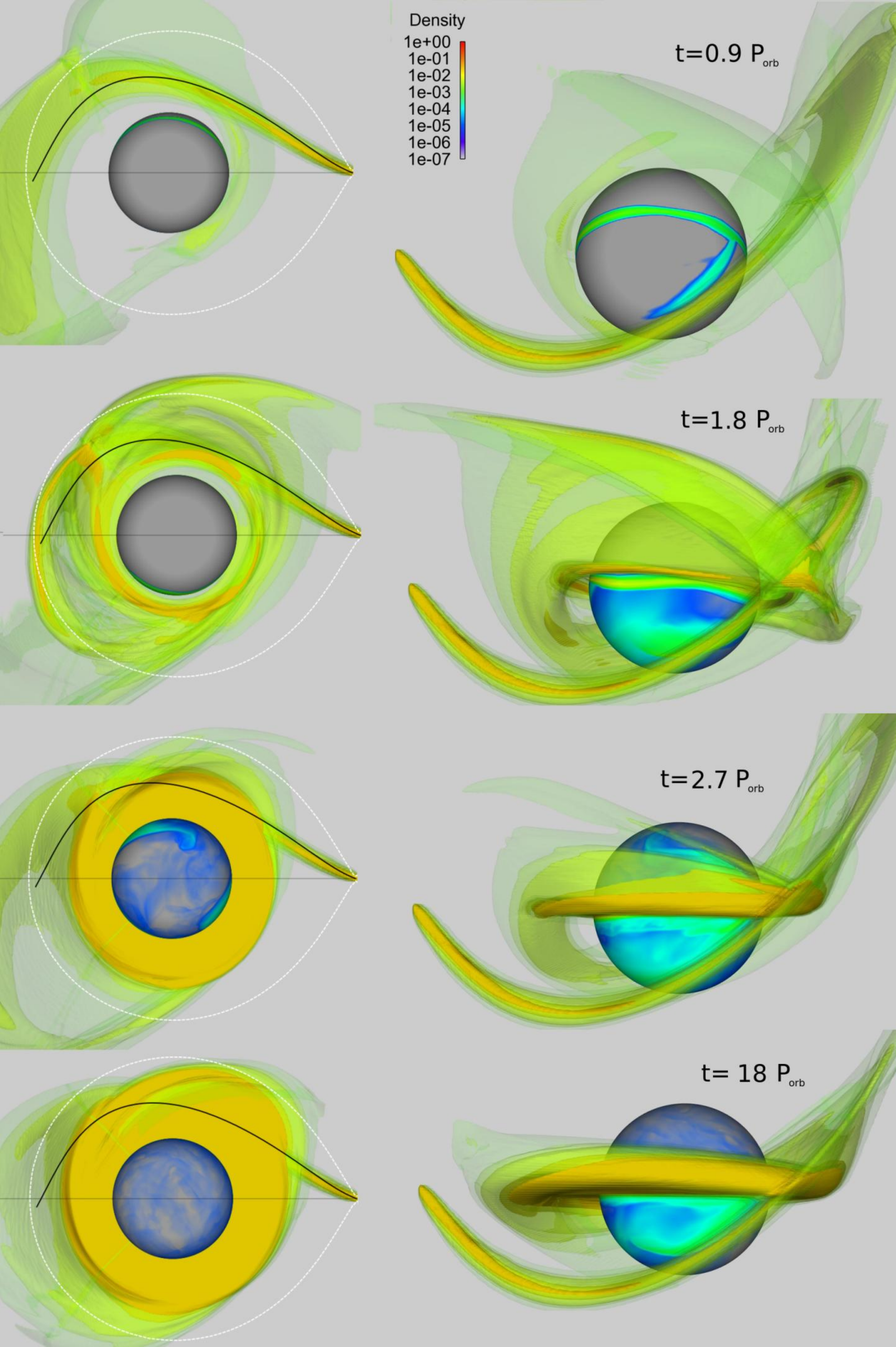}  
\caption{Top and side views of the primary star and disc for a stream boosted to Mach 30 and deflected $45^{\circ}$ below the orbital plane. The density scale is normalized to 1 at L1. Density isosurfaces are taken at $\rho = 1 \times 10^{-1}, 5 \times 10^{-2}, 1 \times 10^{-2}, 5 \times 10^{-3}, 1 \times 10^{-3}$ with the transparency increased by a factor of 0.5 for each isosurface. Due to asymmetry about the orbital plane, density isosurfaces are depicted in the top view, as opposed to a slice along the plane.}
\label{fig:tilt_topside}
\end{minipage}
\end{figure*}

To simulate a system with a deflected and a boosted velocity, we modified the stream parameters at the zones corresponding to L1 by directing the stream velocity $45^{\circ}$ below the orbital plane and boosting it to thirty times the local sound speed; this is comparable to gas flows within magnetic loops. These values were chosen to maximize the likelihood that a tilted disc would form. In the early stages of evolution (Figure \ref{fig:tilt_topside}), the stream falls along the predicted ballistic trajectory, which extends below the orbital plane to just below the bottom of the accreting star. As it nears the primary, it is pulled back toward the orbital plane and wraps around the accretor. The boosted velocity is large enough to cause the stream to miss the surface of the star, resulting in the absence of a star-stream collision and a gas stream that extends above the orbital plane to nearly $2 a$. The stream then circles the accretor and interferes with the gas stream from the secondary as an accretion disc begins to form near the orbital plane. The flow during the first few orbital periods displays a large degree of out-of-plane flow as the system approaches a steady state. At $2 P_{orb}$ the density within the disc is boosted, inducing some minor radial oscillations as in the undeflected case. As the disc gains mass, the stream-disc interaction continues to drive gas upward above the orbital plane. The system stabilizes over the next $15 P_{orb}$, after which the accretion disc remains present along with a large plume of gas deflected above the orbital plane by the stream-disc interaction.

Investigating the morphology of the accretion disc shows some radial enlargement and warping due to the deflected stream. Averaging the density in the central plane over $\phi$ (Figure \ref{fig:density_v_radius}) shows that the density falls off more slowly than in the undeflected case, reaching $10^{-3}\rho_{L1}$ at $0.5a$. The inner density of the disc remains on the same order of magnitude as in the undeflected case, and is slightly larger in radius. The top view of Figure \ref{fig:tilt_topside} at $18 P_{orb}$ shows that the disc is also asymmetric, with a slightly enlarged portion in the upper right portion of the disc. This large density region originates as the disc forms and propagates around the disc with time. It does not appear to originate due to the artificial density boost that occurs at $2 P_{orb}$. The vertical extent of the disc is similar to that of the undeflected case, neglecting slight enlargement above and below the orbital plane due to warping that occurs in the outer edges of the disc. The majority of the disc still lies in the orbital plane, but with lower density gas ($\sim 0.1 \rho_{L1}$) deflected above the plane due to the stream-disc interaction. This gas is pulled back into the plane, where further dissipation of angular momentum due to interactions with the disc causes a portion of the material to join the disc and eventually accrete. 

The deflected stream begins at a speed of $285$ km s$^{-1}$ (Mach 30). As the circumprimary material begins to build up into a disc, gas is blown upward in the collision with z-velocities of up to $210$ km s$^{-1}$. Within the disc we observe a circular orbit with velocities that remain sub-Keplerian in the same range as the undeflected case ($240-440$ km s$^{-1}$ for the high-density portion, $150-250$ km s$^{-1}$ for the low-density portion). We also measure z-velocities on the order of $\pm $ 10 km s$^{-1}$ due to the warping and tilting in the disc. 

Examining the tilt within the disc as a function of radius after $18 P_{orb}$ has elapsed (Figure \ref{fig:tilt_v_radius}) shows that the warping within the disc is more pronounced than in the undeflected stream. The inner portion of the disc is tilted at $\sim 1.5^{\circ}$ near the the surface of the primary, and increases to $\sim 3^{\circ}$ at $0.35a$. Warping is indicated by the variations in tilt angle with increasing radius. The tilt continues to increase at radii greater than $0.35a$, mostly due to the high z-velocities in the material blown upward in the stream-disc interaction. The tilt angle calculated for this low-density gas reaches values as high as $50^{\circ}$ at the outer edge of the disc. All the tilt measurements for this timestep result in positive values, indicating that the disc is tilted consistently in the $+x$ direction (toward the secondary). 

The orientation of the disc changes noticeably over time when we deflect and boost the stream. The plot of $J_y$ vs. $J_x$ (Figure \ref{fig:orientation}) shows a much wider spread of values than that of the undeflected stream. The fraction of angular momentum that is distributed in the $x$ and $y$ components remains between 2 and 3 percent at all times, and grows to 5 percent when the disc is oriented toward the donor star. When this occurs, the effects of the stream's impact on the disc are enhanced due to the upward force of the stream on the already upwardly-moving gas in the disc. As in the undeflected case, the apparent precession is due to measurements from within the corotating reference frame.

\subsection{Emissivity}

Additional information regarding the out-of-plane velocities and accretion features are more easily visible if we mimic the tomographic results with a pseudo-emissivity plot in velocity space. This was first done for Algol systems by \citet{richards98}. In our simulations, emissivity scales as $\rho^2,$ and lacks any dependence on the temperature or ionization state of the gas. We slice our velocity space plots into slabs along $v_z$, sum the emissivity along $v_z$ within each slab, and scale the emissivity to the maximum value observed in the slab. This maximum value is also listed in terms of the maximum emissivity on the entire grid. The velocities measured in our hydrodynamic simulations take place in a corotating frame, and we transform these into an inertial frame for the sake of comparison with tomographic data. Since we use the inner boundary to represent the primary and we have removed the secondary from the grid, emission features associated with stellar activity occuring on the surface of either star (e.g. circumprimary emission, chromospheric emission from the secondary) are absent from the plot. As a result, the plots appear more sparse than the original tomographic data presented in \citet{agafonov09} and \citet{richards10}. We also note that the strong density dependence causes the emissivity in the central plane to appear many orders of magnitude greater than that of other slices.

\begin{figure*}
\begin{minipage}{126mm}
\includegraphics[width=126mm]{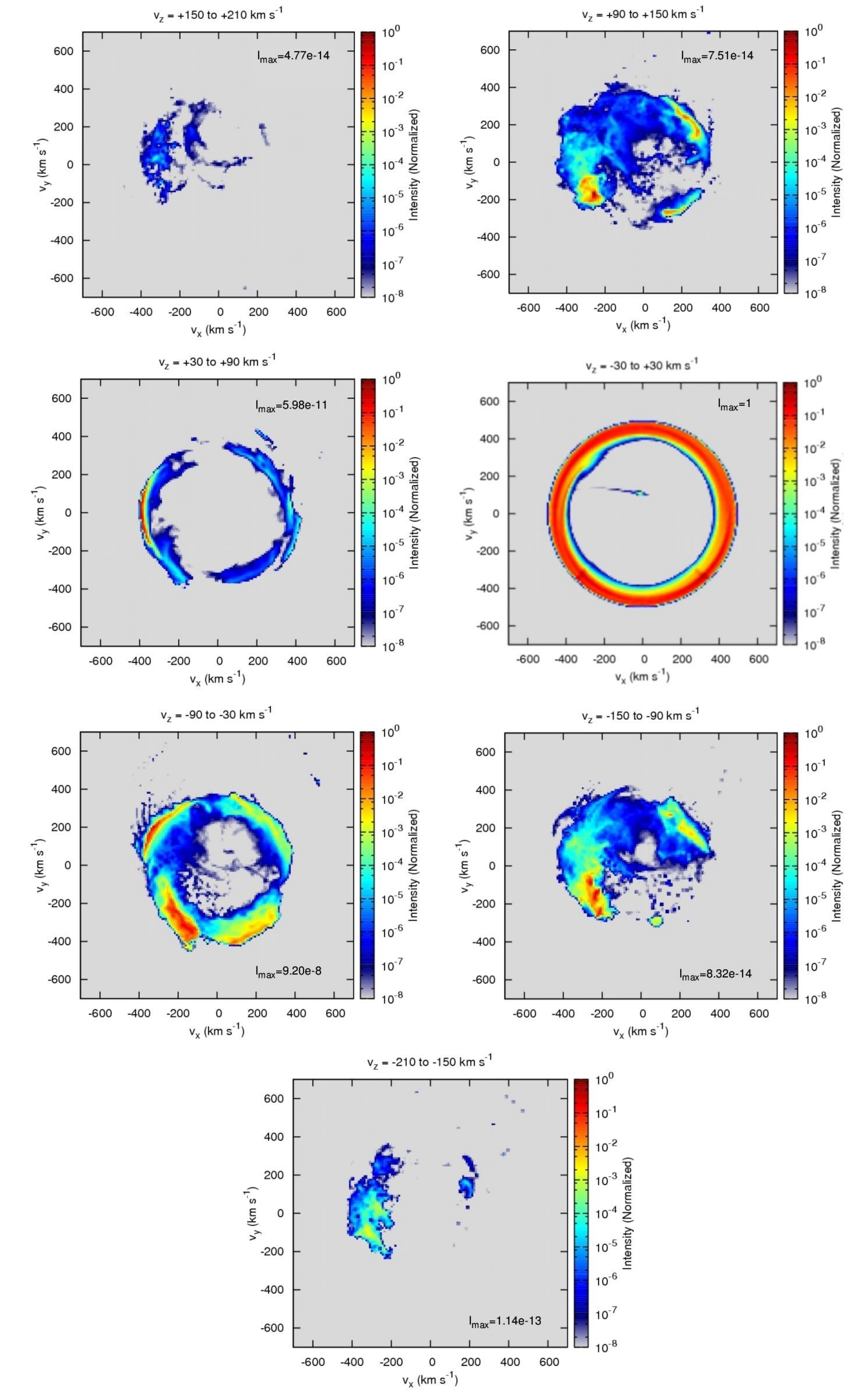}  
\caption{Emissivity for RS Vul with an undeflected stream after $18 P_{orb}$. The emission intensity on each plot is normalized to the maximum within the individual slice. The value of each maximum is also listed in terms of the maximum intensity on the entire grid.}
\label{fig:halpha_nt}
\end{minipage}
\end{figure*}

The emissivity plots for the undeflected stream (Figure \ref{fig:halpha_nt}) clearly show the presence of the disc in the central slice ($v_z = -30$ to $30$ km s$^{-1}$). Since the inner portion of the disc circles the primary with a higher velocity than the outer portion, the disc is turned inside-out on the velocity space plot. The bulk of the disc possesses velocities that fall within the Keplerian predictions (400-515 km s$^{-1}$), although we also measure sub-Keplerian velocities as small as 370 km s$^{-1}$. The accretion stream is also visible in the central slice, beginning near $v_x = 0$ km s$^{-1}$ and extending further in the -x direction as it gains speed and falls along the binary axis. The slightly stronger emissivity region located along the inside region of the``ring'' at roughly $(v_x, v_y)$ = ($-250,250$) km s$^{-1}$ corresponds to the stream-disc interaction. Emission from this feature extends up to $v_z=30$ km s$^{-1}$. We also note that the emissivity is slightly enhanced near ($\pm 350,-350$) km s$^{-1}$ due to the overlap between the yin and yang grids.

Some emission is present in the slab corresponding to $v_z$ = 30 to 90 km s$^{-1}$, and is located primarily within the locus of the accretion disc. The intensities of this emission are seven orders of magnitude below the maximum intensity on the grid, and distinct emission features are not immediately apparent in this slice. For gas wtih downward velocities of $v_z$ = $-30$ to $-90$ km s$^{-1}$, some emission is present in the locus of the disc again, though at intensities ten orders of magnitude below the maximum grid intensity. For velocities greater than $v_z$ = $\pm$90 km s$^{-1}$, the emission lacks any distinct features and occurs at intensities on the order of $10^{-13}-10^{-14}$ of the maximum intensity. 

\begin{figure*}
\begin{minipage}{126mm}
\includegraphics[width=126mm]{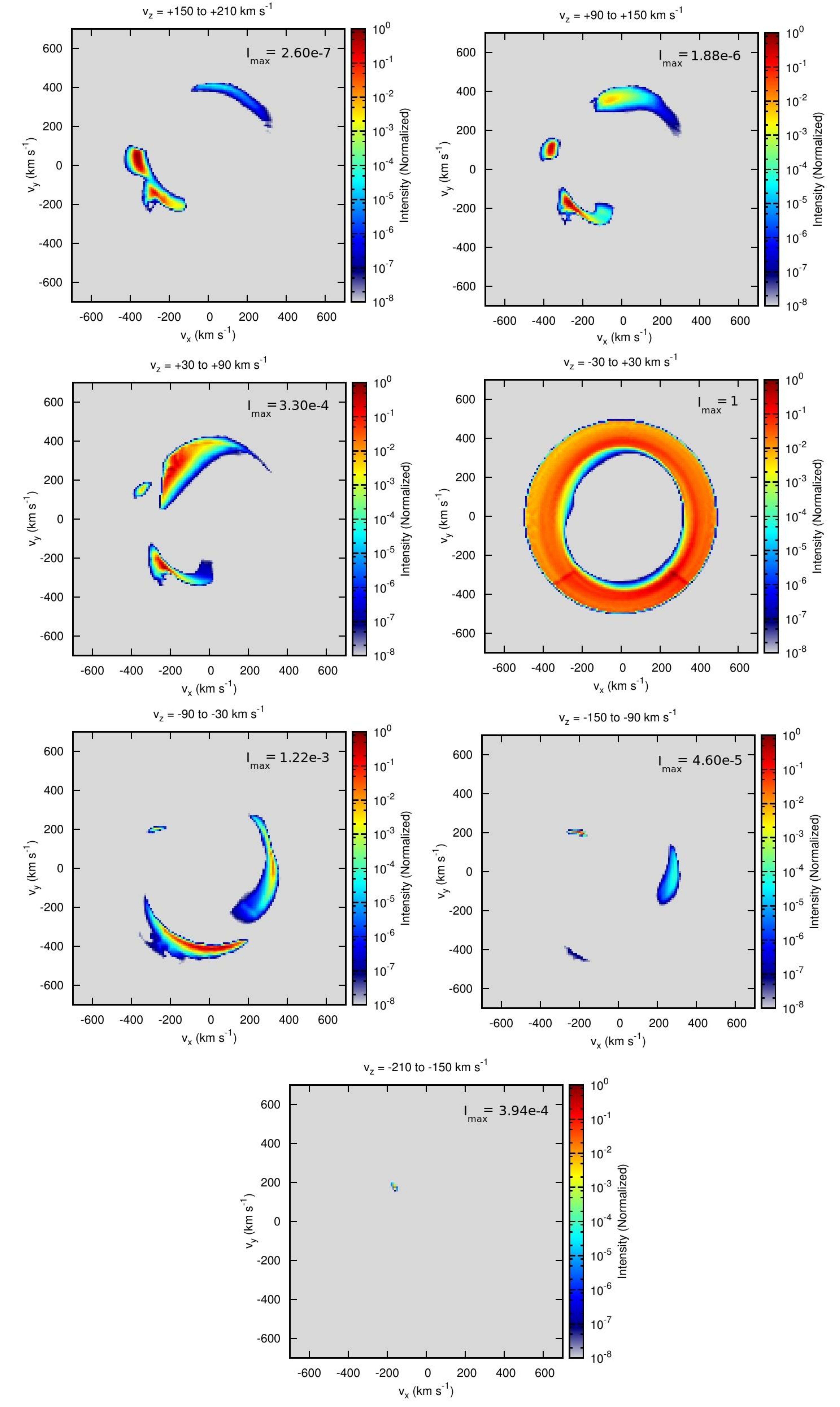}  
\caption{Emissivity for RS Vul with an accretion stream deflected below the orbital plane by $45^{\circ}$ and boosted to Mach 30 at $18 P_{orb}$. As in the previous figure, the emission intensity is normalized to the maximum value within each individual slice, and the value of each maximum is listed in terms of the maximum intensity on the entire grid.}
\label{fig:halpha_tilt}
\end{minipage}
\end{figure*}

The emissivity plots of the tilted disc (Figure \ref{fig:halpha_tilt}) show much higher emissivity measurements in slices above and below $v_z=0$. The disc is again the dominant feature in the central slice, and is larger in radius than in the undeflected case, as expected from the density profiles. This results in an emissivity region that extends inward to a slightly greater extent. The majority of the velocities lie within the Keplerian predictions ($375-515$ km s$^{-1}$), with some gas possessing velocities as low as 350 km s$^{-1}$. Since the stream has been deflected below the orbital plane, it appears in the upper left quadrant of the $-v_z$ slices. The stream-disc interaction is visible in the slices corresponding to positive z-velocities, and appears most prominently near $(v_x, v_y)$ = ($-450,150$) km s$^{-1}$ in the slice of $v_z = 30$ to $90$ km s$^{-1}$. The high $v_z$ material on the far side of the disc is also apparent in the emissivity plots. As it rises and wraps around the secondary (with $+v_z$), it can be seen in the lower left quadrant. As it falls back towards the orbital plane ($-v_z$), it shifts toward the lower right quadrant, indicating motion back toward L1. The gas collides with the orbital plane again, and a portion is driven below the orbital plane once again. This is visible in the $+v_z$ slices as a feature in the upper right quadrant.

As a whole, the emissivity for the deflected and boosted stream reveals values above and below $v_z=0$ with intensities much closer to the values measured in the central slice. This indicates a much stronger out-of-plane flow produced by the deflected and boosted stream. It should also be noted that the maximum intensity on the grid is nearly five times larger for the deflected and boosted stream than for the undeflected stream. This is to be expected, since the densities within the disc are slightly larger when the stream was deflected, and any changes in density will appear in the emissivity as $\rho^2$. Examining the emissivity in the early stages of our run ($2 P_{orb}$, just before the density is boosted) shows the intensity more uniformly distributed throughout the $v_z$ slices. No notable differences in the structure of the emissivity features are seen.

\section{Analysis}

The main features visible on our emissivity plots for both the undeflected and deflected streams are the disc emission, accretion stream, and stream-disc interaction. In the undeflected case we also measure non-localized ``clouds'' with z-velocities up to $\pm 210$ km s$^{-1}$, but with magnitudes smaller than the emission features located in the central plane by a minimum of eight orders of magnitude. For the deflected stream, the main sources of emissivity are the disc and the out-of-plane gas located on the far side of the secondary (extending from $v_z=+210$ km s$^{-1}$ to $v_z=-150$ km s$^{-1}$). The disc structure is similar to that of the undeflected case, but displays some radial enlargement and slight warping. The out-of-plane gas appears with much stronger intensities relative to the gas in the orbital plane, and measures $10^{-3}-10^{-7}$, as opposed to $10^{-8}-10^{-14}$.

The early stages of our deflected stream simulation are comparable to the tomographic data of RS Vul \citep{richards10}. The emission seen in the tomographic data of RS Vul is not confined to the Keplerian locus and is found primarily along the stream, near the mass donor and accretor, and in several other localized high-velocity regions. In our simulations, prior to the formation of the accretion disc (i.e. before the density boost occurs at $\sim 2 P_{orb}$) we observe a similar structure. The stream is visible in slices with negative velocities as low as $v_z=-210$ km s$^{-1}$, and the high-velocity features corresponding to the stream material wrapping around the primary are visible from $v_z = -150$ km s$^{-1}$ to $v_z = +210$ km s$^{-1}$. As in the tomographic data, the star-stream interaction is not visible, since the deflected stream misses the star entirely. Emission from the Keplerian locus is non-uniform and occurs at intensities lower than those corresponding to the stream by an order of magnitude.

Despite similarities with the morphology of our simulations, the tomographic data from RS Vul show a more feature-rich display, with out-of-plane velocities as high as $v_z = \pm 480$ km s$^{-1}$. These are much larger than the $\pm 210$ km s$^{-1}$ maximum we measure in our emissivity plots, but the magnitude of the tomogram velocities may be somewhat exaggerated due to the data's relatively low resolution in the z-direction. Another important difference from our model is that the intensities of the emission features in the tomograms remain on the order of those in the central plane even at very high values of $v_z$. The primary explanation for our lower intensity measurements is that the emission of low-density regions is suppressed by our use of an isothermal gas. Since the emissivity calculation depends strictly on density in the isothermal case, the magnitude of the emission in the simulation data appears smaller than it otherwise would in regions containing hot low-density gas. Since we have neglected the influence on the ionization state and temperature of the gas in our emissivity plots, we expect our intensities to fall drastically in the lower-density portion of the flow. In the early stages of the run, this results in the maximum intensities occurring at the location of the stream. The intensity elsewhere is smaller by a factor of $10^{-2}-10^{-4}$.

The late stages of our simulation display the accretion disc in a steady-state with a  morphology more akin to the tomographic data of U CrB in its disc mode \citep{agafonov09}. This tomographic data shows more concentrated emission in the Keplerian locus than the tomograms of RS Vul, as well as both a star-stream and stream-disc interaction region. Our late-stage emissivity plots are also dominated by emission from the Keplerian locus due to the presence of a disc, but the material driven away from the orbital plane in the stream-disc interaction remains visible. Due to the strong density dependence of the emissivity, the disc dominates the emissivity plots. Intensities elsewhere range from $10^{-3} - 10^{-7}$ below the maximum measurement on the central slice.

Our measurements of the tilt angle of the disc show that deflecting and boosting the stream introduces warping of up to $3^{\circ}$, but no significant tilt with respect to the orbital plane. Alternately, the lower-density gas blown upward above the orbital plane does display significantly larger tilt that measures up to $50^{\circ}$ at the outer edge of the disc. Obtaining a tilt measurement from the tomographic data for comparison is difficult, particularly for RS Vul, which does not display a well-defined accretion disc. In U CrB, z-velocities of $\pm 420$ km s$^{-1}$ were observed within the predicted locus of the Keplerian accretion disc. A tilted disc was inferred from the observation that emission in tomogram slices corresponding to negative z-velocities was stronger on the left side of the $v_y-v_x$ velocity-space plots, while positive z-velocity slices showed stronger emission in the lower-right side of the $v_y-v_x$ plot. Quantifying this tilt by assuming a z-velocity of $420$ km s$^{-1}$ and Keplerian velocities at the surface of the primary gives a tilt angle of $37.2^{\circ}$. Performing the same calculation at the outer Keplerian orbit produces a tilt of $56.1^{\circ}$, which is comparable to the tilt measured in our simulations for the plume of gas on the far side of the accretor. It should also be noted that the $v_z$ values are uncorrected for the lower resolution in the $v_z$ direction. \citet{agafonov06} claim the correction factor will be small, but any correction factor will serve to reduce the predicted tilt. Assuming a factor as large as $0.5$ would reduce the maximum z-velocities to $v_z = \pm 240$ km s$^{-1}$, which is close to the maximum $v_z$ we measure. This correction results in tilts of $20.8^{\circ}$ and $36.6^{\circ}$ at the inner and outer radii. Our simulations indicate that it is unlikely that the disc will be uniformly tilted, since the stream-disc interaction acts primarily on the outer edge of the disc. The tomographic data that depicts asymmetric intensities measured on opposite sides of the accreting star could be interpreted as warping in the disc, as opposed to a uniformly inclined disc. A large part of the out-of-plane flow in our simulations also occurs within the Keplerian locus, and could contribute to this measurement.

The primary difference between the deflected and undeflected stream in our simulations is that the modified stream enables large amounts of flow above and below the orbital plane. Boosting the stream allows gas to circle the primary without striking its surface, which in turn means the system retains angular momentum that would normally be destroyed in the stream-disc and star-stream collisions. The additional (non-z) angular momentum manifests in the out-of-plane flow, and is eventually dissipated in the accretion disc. The disc's dimensions appear larger and the densities slightly higher due to the absence of the star-stream interaction as an angular momentum dissipation mechanism. The bulk of the disc remains primarily within the orbital plane, although it is more dense and more uniform than the disc shown in the tomographic measurements of U CrB. We can reconcile the differences between the simulated and observed discs by noting that our isothermal approximation results in a gas that is more easily compressible. This will lead to a more compact accretion disc and the tendency for less gas to be blown away in the interaction regions. It is also possibile that the tomographic images of U CrB in a disc mode show the early stages of the disc, in which case the relative intensity of the disc would be lower, and the disc would have a more nonuniform structure.

We see no evidence of a transition between a stream and disc accretion mode in our simulations. This is likely a result of our use of a constant mass transfer rate through L1, and potentially a byproduct of our isothermal equation of state. Examining our emissivity plots at $2P_{orb}$ shows a morphology very similar to that at $18 P_{orb}$, but with the central slice intensity $1-2$ orders of magnitude higher than that on the remainder of the grid. While the disc is present, it is not fully formed, and appears less uniform than late in the run. A reduced rate of mass transfer could offer the disc material a chance to accrete without being replenished by material from the stream. A non-isothermal gas would contribute to this effect by resulting in a less compressible flow that would impede the formation of a stable disc.  During this time, we would expect to observe the system in its stream phase. Since the time necessary for a full disc to form  is large (preliminary runs showed $\dot{M}_{L1} = \dot{M}_{acc}$ was achieved after $40-50 P_{orb}$), variations in the mass transfer rate occurring across smaller timescales would prevent the full formation of a stable disc. Our use of an isothermal gas also increases the compressibility, and subsequently the density of our disc. Relaxing the isothermal approximation could result in a lower-density, less uniform disc.

\section{Conclusions}

Our simulations of RS Vul have shown that boosting the velocity of the accretion stream and deflecting it below the orbital plane at L1 introduces significant changes in the structure of the accretion flow. Using an isothermal equation of state, we simulate the formation of an accretion disc with both an undeflected stream and a stream boosted to Mach 30 and deflected by $45^{\circ}$. In the undeflected case, we measure the formation of an accretion disc that is symmetric about the orbital plane and displays less than one angular grid zone ($0.5^{\circ}$) of warping. Modifying the stream causes some radial enlargement of the disc, along with warping as large as $3^{\circ}$. In addition, a low-density plume of gas is ejected above the orbital plane on the far side of the accreting star with densities on the order of $10^{-2}\rho_{L1}-10^{-3}\rho_{L1}$ and tilt angles as high as $50^{\circ}$.

The intensities we measure in our pseudo-emissivity plots are strongest in the orbital plane due to our use of an isothermal gas, which causes the emissivity to depend solely on $\rho^2$. The increased out of plane flow that occurs due to the deflected stream increases the emissivity above and below the central plane by a minimum of five orders of magnitude. We anticipate that relaxing the isothermal condition would cause higher emissivities to be measured outside of the central plane, as in the original tomographic data. 

Our results are consistent with the hypothesis that stream deflection is responsible for the ``tilted'' out-of-plane material that is measured in the tomograms of U CrB, and also for the extreme out-of-plane flow observed in RS Vul. We have shown that a strong velocity boost and deflection at L1 are sufficient to change the direct-impact nature of the system, which increases the amount of out-of-plane flow drastically. We expect that a variable mass transfer rate and potentially more extreme boosting would only serve to enhance the features observed in the out-of-plane flow.

\section{Acknowledgments}

This research was supported by an allocation of advanced computing resources provided by the National Science Foundation. The computations were performed on Kraken at the National Institute for Computational Sciences. This research was also supported in part by a North Carolina Space Grant award. The author would like to thank Dr. John Blondin for his assistance with this work, and the anonymous referee for the helpful suggestions regarding the manuscript.

\clearpage

\end{document}